\documentclass[prd,twocolumn,superscriptaddress,showpacs,preprintnumbers,amsmath,amssymb,APS]{revtex4-1}
\usepackage{bm}
\usepackage[latin1]{inputenc}
\usepackage[spanish,english]{babel}
\usepackage{amsfonts}
\usepackage{amssymb}
\usepackage{graphicx}
\usepackage{color}

\newcommand{\be}{\begin{equation}}
\newcommand{\ee}{\end{equation}}
\newcommand{\bea}{\begin{eqnarray}}
\newcommand{\eea}{\end{eqnarray}}

\begin{document}

\title{{\bf Electromagnetic Duality Anomaly in Curved Spacetimes}}

\author{Ivan Agullo}\email{agullo@lsu.edu}\affiliation{Department of Physics and Astronomy, Louisiana State University, Baton Rouge, LA 70803-4001;}
\author{Adrian del Rio}\email{adrian.rio@uv.es}
\affiliation{Departamento de Fisica Teorica and IFIC, Centro Mixto Universidad de Valencia-CSIC. Facultad de Fisica, Universidad de Valencia, Burjassot-46100, Valencia, Spain.}\affiliation{Department of Physics and Astronomy, Louisiana State University, Baton Rouge, LA 70803-4001;}
\author{Jose Navarro-Salas}\email{jnavarro@ific.uv.es}
\affiliation{Departamento de Fisica Teorica and IFIC, Centro Mixto Universidad de Valencia-CSIC. Facultad de Fisica, Universidad de Valencia, Burjassot-46100, Valencia, Spain.}

\date{\today}

\begin{abstract}


The source-free Maxwell action is invariant under electric-magnetic duality rotations in arbitrary spacetimes. This leads to a  conserved classical  Noether charge. We show that this conservation law is broken at the quantum level in the presence of a background classical gravitational field with a nontrivial Chern-Pontryagin invariant,  
in parallel  with the  chiral anomaly for massless Dirac fermions.
 Among the physical consequences, the net polarization of the quantum electromagnetic field  is not conserved.

\end{abstract}


\maketitle


{\bf  1. Introduction. }
It has long been known that the {\em source-free} Maxwell equations in four dimensions 
are manifestly invariant  under duality rotations of the electromagnetic field $F_{\mu\nu} \to F_{\mu\nu} \cos \theta + {^{\star}F}_{\mu\nu}\sin \theta $, where ${^{\star}F}_{\mu\nu}$  is the dual strength tensor. It was proven in Ref. \cite{deser-teitelboim} that this transformation is indeed a symmetry of the action---at the level of the basic  dynamical variables $\vec A$, and for an arbitrary spacetime---and the associated conserved charge was identified.  
%
This symmetry extends to the quantum theory in Minkowski spacetime. The goal of this Letter is to analyze whether the duality invariance persists in quantum field theory in curved spacetimes or, as for the  chiral invariance of massless fermions, the presence of spacetime curvature  induces an anomaly. 


 
 If the symmetry exists and  leaves the vacuum state invariant, vacuum expectation values of operators that reverse sign under a discrete duality transformation
, such as $ {F}_{\mu\nu}{F}^{\mu\nu}(x)=2\left[\vec B^2(x) - \vec E^2(x)\right]$, must vanish. However, it has been found in Ref. \cite{ALN14} using adiabatic renormalization that this is not the case for a spatially flat Friedmann-Lemaitre-Robertson-Walker (FLRW) spacetime,
   %
\be  \langle {F}_{\mu\nu}{F}^{\mu\nu} \rangle=
 \frac{1}{ 480 \pi^2}\left(-9 R_{\alpha\beta}R^{\alpha\beta}+\frac{23}{6}R^2  +4\Box R\right)  \, , \ee
 where $R_{\alpha\beta}$ is the Ricci tensor and $R$ is its trace. This result signals a breaking of the duality symmetry. On the other hand, the same approach produces a vanishing  value of    $\langle {F}_{\mu\nu}{^{\star}F}^{\mu\nu}\rangle = -4\langle \vec E \cdot \vec B \rangle$ in FLRW spacetimes. Given its  pseudoscalar character, this quantity is expected to be proportional to the  Chern-Pontryagin  invariant density $ R_{\mu \nu \lambda \sigma} {^{\star}R} ^{\mu \nu \lambda \sigma}$, where ${^{\star}R} ^{\mu \nu \lambda \sigma}= 1/(2 \sqrt{-g})\epsilon^{\mu\nu\alpha\beta } R_{\alpha \beta}^{\ \ \ \lambda \sigma} $, which vanishes in FLRW backgrounds. This  was  indeed worked out  in Ref. \cite{DKZ87}, finding that  
 \be \langle {F}_{\mu\nu}{^{\star}F}^{\mu\nu}\rangle = \frac{1}{48 \pi^2}  R_{\alpha \beta \lambda \sigma} {^{\star}R}^{\alpha \beta  \lambda \sigma} \, . \label{rusos}\ee

 Although these results  are  suggestive, to establish the existence of an anomaly one needs to go a step further and analyze the extension of the classical conservation law to the quantum theory.  This was the strategy followed to prove the existence of  the chiral anomaly for spin-$1/2$ fermions interacting with an external electromagnetic  \cite{ABJ} or a gravitational field \cite{Kimura} in the late 1960s. The purpose of this  Letter is to build a similar formalism for the electromagnetic field. Even though the gauge freedom adds new difficulties, 
the analysis brings out an interesting formal relation between the electromagnetic  and fermionic dynamics that happens to be of great utility. We show that Maxwell equations in radiation gauge can be rewritten as spin 1 Dirac-type equations $\beta^{\mu}\nabla_\mu \Psi=0$, where $\Psi$ is a two-component object made of the potentials of the  self- and antiself- dual parts of the electromagnetic  field (these components describe right and left circularly polarized waves, respectively), and the matrices $\beta^{\mu}$ are   spin 1 analogs of the familiar $\gamma^{\mu}$ matrices for spin-$1/2$ fermions. Duality rotations are then generated by $\beta_5$, which is defined in the standard way (see below). The extensive theoretical machinery developed to derive the fermionic chiral anomaly can then be extended to the electromagnetic case. In particular, following the well known Fujikawa method \cite{Fujikawa, Fbook}, we show that the duality anomaly originates in the failure of the measure of the path integral to respect the  symmetry of the action. In the rest of the Letter we spell out the details of the analysis and summarize the interesting relation with other mathematical structures and the physical consequences of the anomaly.  

We  follow the convention $\epsilon^{0123}=1$ and   metric signature $(+, -, -, -)$.\\
  
{\bf 2. Duality symmetry and Noether charge}.
A detailed analysis of the duality symmetry of the classical, source-free Maxwell theory was presented in Ref. \cite{deser-teitelboim} (see Ref. \cite{Calkin} for an earlier work), and the reader is referred to these references for details.  At the level of the electromagnetic potential, duality rotations are implemented by the transformation  $\delta A_{\mu}=\theta \, Z_{\mu}$, with $\theta$ an infinitesimal parameter. $Z_{\mu}$ is a vector field that, on shell, must satisfy $\nabla_{\mu} Z_{\nu} - \nabla_{\nu} Z_{\mu} = {^{\star}F}_{\mu\nu}$, and can be understood as a nonlocal function of  the basic variables $A_{\mu}$ \cite{z0}. 
By taking the exterior derivative, one can see that the transformation above reduces to the more familiar form $\delta F_{\mu \nu}=\theta \ {^{\star}F}_{\mu\nu}$ on shell. The associated conserved current can be  obtained from the Lagrangian density, and reads 
 \be \label{jND} j^{\mu}_D= \frac{1}{2}(A_{\nu}{^{\star}F}^{\mu\nu} -I^{\mu\nu} Z_{\nu}) \ , \ee
where $I_{\mu\nu}=2F_{\mu\nu}+\sqrt{-g}\epsilon_{\mu\nu\alpha\beta} \nabla^{\alpha}Z^{\beta}$, which satisfies $I_{\mu\nu}=F_{\mu\nu}$ on shell. This current is gauge dependent and nonlocal in space. However, the integral on a spatial Cauchy hypersurface of its ``zero component'' produces a well defined, gauge invariant {\em conserved charge} $Q_D$, 
 which physically accounts for the difference in amplitude between left and right polarized components of the electromagnetic radiation. This quantity is often called the optical helicity,  and it is the Noether generator of duality transformations. 
 
 Note that the first term in Eq. (3) is  proportional to the Pauli-Ljubanski vector $K^{\mu}\equiv -A_{\nu}{^{\star}F}^{\mu\nu}$ used in Ref. [3] to compute $\langle {F}_{\mu\nu}{^{\star}F}^{\mu\nu}\rangle =-2\left< \nabla_{\mu} K^{\mu}\right>$. This vector is not conserved, already at the classical level. Moreover, the spatial integral of $K^0$ is related to the so-called magnetic helicity \cite{Berger1999}, which does not generate electromagnetic duality transformations. The second term in Eq. (3), $-\frac{1}{2}I^{\mu\nu}Z_{\nu}$, is needed for the current  to be conserved in the classical theory and the associated charge to generate duality rotations. The  evaluation of  $\langle \nabla_{\mu}j^{\mu}_D \rangle$, which turns out to be  equivalent to $ -\left<(\nabla_{\mu}F^{\mu\nu})Z_{\nu}\right>$, requires us to deal with an operator different from  $ F_{\mu\nu}{^{\star}F^{\mu\nu}}$ and $ F_{\mu\nu}{F^{\mu\nu}}$. Consequently,  a different calculation needs to be elaborated in order to analyze the electromagnetic duality  in the quantum theory. \\


{\bf 3. Weyl-type  representation of Maxwell's equations}. 
 Before moving to the analysis of the above conservation law in the quantum theory, we rewrite Maxwell's equations in a convenient  form for our purposes.  We first describe the formalism in Minkowski spacetime and then generalize it to other geometries. 
 
 Maxwell's equations in the absence of charges and currents  in Minkowski spacetime {\em decouple} when written in terms of the complex fields $\vec H_{\pm}\equiv \frac{1}{2}[ \vec E\pm i\vec B]$ with $\vec E$ and $\vec B$ the electric and magnetic fields, and take the form
\bea \label{eqsH}
 \vec \nabla \times \vec H_{\pm}  = \pm \, i \frac{\partial}{\partial t} \vec H_{\pm}\ , \hspace{0.3cm} \vec \nabla\cdot \vec H_{\pm}=0 \label{Maxwellequationsforthefield}.
 \eea
 Using the familiar transformation properties of $\vec E$ and $\vec B$ under the Lorentz group, it is straightforward to show that $\vec H_{+}$ and $\vec H_{-}$ transform according to the $(1,0)$ and $(0,1)$ representations, respectively. Under a duality transformation $\vec{E}\to \cos{\theta} \vec{E}+\sin{\theta} \vec{B}$, $\vec{B}\to -\sin \theta \vec{E}+\cos \theta \vec{B}$, we have $\vec H_{\pm}\to e^{\mp i \theta} \vec{H}_{\pm}$. Hence, $\vec{H}_+$ and $\vec{H}_-$ are the self- and antiself- dual parts of the electromagnetic field, respectively. Interestingly, duality rotations on $ \vec{H}_{\pm}$ resemble conventional chiral rotations in the Dirac theory. Moreover, Eqs. (\ref{eqsH}) can be rewritten as 
 \be \label{alphaMaxwell} (\alpha^{a})^{b}_{\hspace{0.15cm}i} \partial_{a} H^i_+=0 \ , \hspace{1cm}    (\bar \alpha^{a})^b_{\hspace{0.15cm}i} \partial_{a} H^i_-=0   \ , \ee
 (the bar denotes complex conjugation), where the spacetime indices $a,b$ run from 0 to 3, and the internal index $i$ runs from  1 to 3 [note that $\vec H_{+}$ belongs to a three-dimensional complex space associated with the $(1,0)$ Lorentz representation; and analogously $\vec H_{-}$ to the $(0,1)$ one].  Equations similar to Eq. (\ref{alphaMaxwell}) were also written in Refs. \cite{Weinberg64, Dowker}. The components of the $(\alpha^{a})^{b}_{\hspace{0.15cm}i}$  matrices can be extracted from Eq. (\ref{eqsH}), and it can be checked that they satisfy the following properties  
 \bea \label{propalpha}  \alpha^{(a}\alpha^{b)} \equiv \frac{1}{2}\left[(\alpha^{a})^{c}_{\ i} (\alpha^{b})_{c}^{\ j} + (\alpha^{b})^{c}_{\ i} (\alpha^{a})_{c }^{\ j}\right]   &=& \eta^{ab} \delta_{i}^{j}  \, ,  \\
\alpha^{[a}\alpha^{b]} \equiv \frac{1}{2}\left[ (\alpha^{a})^{c}_{\ i} (\alpha^b)_{c}^{\ j} - (\alpha^{b})^{c}_{\ i} (\alpha^{a})_{c }^{\ j} \right]  &=& -2\, \left[^{+}\Sigma^{a b}\right]_{i}^{\ j} \, , \nonumber\eea 
%
where $^{+}\Sigma^{a b}$ is the generator of the $(1,0)$ representation of the Lorentz group, and $\eta_{ab}$ is the Minkowski metric.  Note the analogy with the properties of the $\sigma^{\mu}=(I,\vec{\sigma})$ matrices that appear in the Weyl equations ($\vec{\sigma}$ are the Pauli matrices) for massless spin-$1/2$ fermions. 

 Equations (\ref{alphaMaxwell}) are equivalent to the more conventional, manifestly Lorentz-invariant equations $\partial_a ^{\pm}F^{ab}=0$, due to the fact that the matrices  $(\alpha^{a})^{b}_{\hspace{0.15cm} i}$ provide an isomorphism between $\vec{H}_{+}$ and the self-dual part of $F^{ab}$,
 $^{+}F^{ab}=  (\alpha^{a})^{b}_{\hspace{0.15cm} i} H^i_{+}$, where $^{\pm}F^{ab}\equiv \frac{1}{2}( F^{ab}\pm i \, {^{\star}F^{ab}})$. Similarly, $^{-}F^{ab}=  (\bar \alpha^{a})^{b}_{\hspace{0.15cm} i} H^i_{-}$.
 Therefore, the $(\alpha^{a})^{b}_{\hspace{0.15cm} i}$ matrices can also be thought of as the analog of the $\sigma^{a}_{AA'}$ or $\alpha^{a}_{\alpha \dot\alpha}$ maps that relate spinors and spacetime vectors \cite{Geroch}.

 Given the divergenceless condition in Eq. (\ref{Maxwellequationsforthefield}), we can now introduce potentials $\vec A_{\pm}$ for $\vec H_{\pm}$: $\vec H_{\pm}= \pm i\, \vec \nabla \times \vec A_{\pm}$. In order to isolate the dynamical degrees of freedom we work in the radiation gauge, $\vec \nabla \cdot \vec A_{\pm}=0$. With this choice, Eqs. (\ref{eqsH})   translate to first-order differential equations for $\vec A_{\pm}$
 \bea
   \vec \nabla \times \vec A_{\pm}  =\pm  i \frac{\partial}{\partial t} \vec A_{\pm}\ , \hspace{0.3cm} \vec \nabla\cdot \vec A_{\pm}=0, \label{MaxwellA}
 \eea
which turn out to have the same form as Eq. (\ref{eqsH}). Therefore, they can also be   written as Weyl-type equations 
 \be \label{alphaMaxwellA} (\alpha^{a})^{b}_{\hspace{0.15cm}i} \partial_{a} A^i_+=0 \ , \hspace{1cm}    (\bar \alpha^{a})^{b}_{\hspace{0.15cm}i} \partial_{a} A^i_-=0   \ . \ee
%
Notice that first-order differential equations are obtained at the expense of working with complex fields $\vec{A}_{\pm}$, and therefore duplicating the number of independent variables. It is not difficult to see that Eqs. (\ref{MaxwellA}) are equivalent to Hamilton's equations for the canonical formulation of Maxwell's theory if we split them into real and imaginary parts  \cite{footnote}.
 The familiar second order differential equations $\Box \vec{A}_{\pm}=0$ arise from Eq. (\ref{alphaMaxwellA}) by acting with the operator   $(\alpha^{c})_{b}^{\  j} \partial_{c}$ in the first equation and with $(\bar \alpha^{c})_{b}^{\ j} \partial_{c}$  in the second one, and then using the properties written in Eq. (\ref{propalpha}).

The generalization to curved spacetimes follows the same procedure as for the Dirac case. Namely, 
Eqs. (\ref{alphaMaxwell}) for the fields translate to 
\be \label{alphaMaxwellcurved} (\alpha^{\mu})^{\nu}_{\hspace{0.15cm}i} \nabla_{\mu} H^i_+=0 \ , \hspace{1cm}    (\bar \alpha^{\mu})^{\nu}_{\hspace{0.15cm}i} \nabla_{\mu} H^i_-=0   \ , \ee
and similarly for the potentials
\be \label{alphaMaxwellcurvedA} (\alpha^{\mu})^{\nu}_{\hspace{0.15cm}i} \nabla_{\mu} A^i_+=0 \ , \hspace{1cm}    (\bar \alpha^{\mu})^{\nu}_{\hspace{0.15cm}i} \nabla_{\mu} A^i_-=0   \ , \ee
where the $\alpha$ matrices in curved spacetime are obtained from the flat space ones by using the vierbein formalism
\be (\alpha^{\mu})^{\nu}_{\hspace{0.15cm}i} (x)= e^{\mu}_a(x)\,  e^{\nu}_b(x)\,  (\alpha^{a})^{b}_{\hspace{0.15cm}i}\, .\ee
The equivalence with the familiar Maxwell equations in curved spacetimes, $\nabla_{\mu}\, {^{\pm}F}^{\mu\nu}=0$, is easily shown from Eq. (\ref{alphaMaxwellcurved}) by taking into account that the covariant derivative in the above equations satisfies $\nabla_{\beta} (\alpha^{\mu})^{\nu}_{\hspace{0.15cm}i}=0$.

An even closer analogy with the Dirac equation can be  achieved by combining together the two sets of equations in Eq. (\ref{alphaMaxwellcurvedA})
\bea
\beta^{\mu}\nabla_{\mu}\Psi(x) = 0\ , \label{diracmaxwell} 
\eea
where we have defined \cite{photon}
\bea
\Psi \equiv  \left( {\begin{array}{c}
    A_+^{\ i}  \\
 A_{-\, i}\\
  \end{array} } \right) \ , \hspace{0.6cm}     \beta^{\mu} \equiv i \left( {\begin{array}{cc}
 0 &  (\bar \alpha^{\mu})_{\nu}^{\ i}  \\
 - (\alpha^{\mu})^{\nu}_{\ i} & 0 \\
  \end{array} } \right)      \ .
\eea 
The $\beta^{\mu}$ matrices inherit from the $\alpha^{\nu}$ matrices the following properties
\bea 
\bar \beta^{(\mu} \beta^{\nu)}   & =  &   - g^{\mu\nu}\, \mathbb  I \, ,\\
\bar \beta^{[\mu} \beta^{\nu]}  &  = & 2     \left( {\begin{array}{cc}
 \, ^{+}\Sigma^{\mu \nu}  & 0  \\
 0 & \, ^{-}\Sigma^{\mu\nu}  \\
  \end{array} } \right)  \, ,
\eea
where the parentheses (square brackets) denote symmetrization (antisymmetrization), and $\mathbb  I$ is the identity matrix when acting on  $\Psi$. 
%
Furthermore, we can construct the chiral matrix in the standard way
\bea
 \beta_5 \equiv  i\frac{\sqrt{-g}}{16} \epsilon_{\mu\nu\sigma\rho}
\bar\beta^{\mu}\beta^{\nu} \bar\beta^{\sigma}  \beta^{\rho}  =    \left( {\begin{array}{cc}
  -\mathbb I_{3\times 3}  & 0 \\
 0 &  \mathbb I_{3\times 3} \\
  \end{array} } \right)  \nonumber,
\eea    
%
which can be used to write the duality transformation in the form of a conventional chiral rotation
\bea
\left( {\begin{array}{c}
  A_+^i \\
A_{-\, i} \\
  \end{array} } \right) \rightarrow e^{i\theta \beta_5}\left( {\begin{array}{c}
 A_+^i \\
A_{-\, i} \\
  \end{array} } \right)= \left( {\begin{array}{c}
 e^{-i\theta} A_+^i \\
e^{i\theta}     A_{-\, i} \\
  \end{array} } \right)\ . \label{3.102}
\eea
In analogy with the terminology used for fermions, $A^{i}_{\pm}$ describe right- and left-handed (circularly polarized) radiation.\\


{\bf 4. The quantum anomaly.} 
To explore whether the  classical conservation law extends to the quantum theory, we rely on the well known Fujikawa path integral approach. 
Transition amplitudes for the quantized free electromagnetic field in the radiation gauge can be extracted from the  following path integral \cite{Fbook, constraints}
%
\bea
\left<A_f,t_f |A_i,t_i\right> & =& \int \mathcal{D} X\, \mathcal{D}A_{k}^{(1)}\mathcal{D}A_{k}^{(2)}  e^{i S_M[A] }\ ,
\eea
(the sum over $k$ is understood) where  $A_{k}^{(1,2)}$ represents the two transverse (linear) polarizations of the potential field,  and  $S_M[A]=-\frac{1}{4}\int d^4x \sqrt{-g}F_{\mu\nu}F^{\mu\nu}$ is the Maxwell action.  On the other hand,  $\mathcal{D} X\equiv (-g)^2 \mathcal{D}A_0\det^{1/2}{[D_{\mu}D^{\mu}\delta^{(3)}(x-y)]}$, where $D_{\mu}$ is the spatial covariant derivative. The  radiation gauge fixing is implicitly included in the measure  \cite{Fbook}. 
We now rewrite this expression in terms of $\Psi$. To do this, first we recall the relation between linear and circular polarization 
$
 A^{(1)}_k  =  A^{+}_k+ A^{-}_k $ and $ A^{(2)}_k  =  -i \left[ A^{+}_k- A^{-}_k \right] .
$
 Then the functional measure can be rewritten in the circular basis as $\mathcal{D} A_{k}^{(1)}\mathcal{D} A_k^{(2) } =  \mathcal{D}\bar\Psi \mathcal{D}\Psi  $,
where we have defined $\bar \Psi \equiv \Psi^{\dagger}\beta^0$.
Finally, we arrive at 
\bea
\left<A_f,t_f |A_i,t_i\right> & = & \int  \mathcal{D} X\,  \mathcal{D}\bar\Psi[ A] \mathcal{D}\Psi[ A] e^{i S_M[A] }\ . \label{pathintegral}
\eea
Recall that, despite the notation,  the variables $\Psi$ and $\bar \Psi$ are not Grassmann numbers.

To evaluate the impact of a duality transformation in the path integral (\ref{pathintegral}) we use again Noether's theorem. In quantum field theory and particularly in gauge theories, the second version of the theorem---in which the infinitesimal parameter $\theta$  is promoted to an arbitrary function of space and time subject to appropriate fall-off conditions at infinity---happens to be more convenient (see, e.g., Refs. \cite{Fbook,parker-toms}). The variation of the Maxwell action under a  transformation of the basic dynamical variables $\delta A_{\mu} = \theta (x) \,Z_{\mu}$, with $\delta A_{0}=0$, is
\be \delta S_M = -\int d^4x  \sqrt{-g}\,  \theta (x) \nabla_{\mu} j^{\mu}_D , \ee   
%
where the resulting  $j^{\mu}_D$  agrees with Eq. (\ref{jND}). Note that at the level of the field strength, the transformation implies $\delta F_{\mu\nu}= \theta (x) \, ^{*}F_{\mu\nu} - Z_{\mu}\nabla_{\nu} \theta (x) + Z_{\nu}\nabla_{\mu} \theta (x)$ on shell. This differs from the transformation used in Ref. \cite{reuter}, where it is assumed that $\delta F_{\mu\nu}= \theta (x)\,  ^{*}F_{\mu\nu}$.

Quantum anomalies arise from the noninvariance of the measure in the path integral \cite{Fujikawa, Fbook}. The transformation properties  of the measure are given by the Jacobian $J$,  $\mathcal{D}\bar \Psi' \mathcal{D} \Psi'=J \mathcal{D}\bar \Psi \mathcal{D} \Psi$. Note that the duality rotation leaves  $\mathcal{D} X$ invariant. To evaluate $J$ it is more convenient to move to the Euclidean regime.  Now, the fact that the operator $D=\beta^{\mu}\nabla_{\mu}$ is  Hermitian, guarantees the existence of an orthonormal basis $\Psi_n$ of eigenstates ($D \Psi_n = \lambda_n \Psi_n$) under the  inner product
$(\Psi_n,\Psi_m) \equiv   \int d^4x \sqrt{-g}\Psi_n^{\dagger}\Psi_m=\ell^2 \delta_{nm}$, where $\ell$ is an arbitrary constant with dimensions of length; physical observables are insensitive to its value, so we fix $\ell=1$. With this, the expression for the Jacobian can be derived by expanding the fields  $\Psi$ and $\bar \Psi$ in terms of this complete basis, and reads 
\be \label{jacobian} J=e^{+i 2\sum_{n=0}^{\infty}\int d^4x \sqrt{-g}\theta(x) (\Psi_n^{\dagger}\beta_5\Psi_n)} \ . \ee
From this, the expression for the vacuum expectation value $\left< \nabla_{\mu}j_D^{\mu} \right>$ can be obtained by
recalling that the path integral is independent of the name of variables
\bea
&&\int D\bar\Psi[ A] D\Psi[ A] e^{i S_M[A] }  =  \int D\bar\Psi[ A'] D\Psi[ A'] e^{i S_M[A'] } \nonumber\\
 & = & \int D\bar\Psi[ A] D\Psi[ A] \, J \, e^{i S_M[A]-i \int d^4x\sqrt{-g}\theta(x)\left< \nabla_{\mu}j_D^{\mu} \right>},
\eea
and therefore $\left< \nabla_{\mu}j_D^{\mu} \right>= 2\sum_{n=0}^{\infty} (\Psi_n^{\dagger}\beta_5\Psi_n)$.
The right-hand side of this expression is not well defined (it is ultraviolet divergent) and must be renormalized. We follow a regularization based on the well known heat kernel expansion (see, e.g., Refs. \cite{parker-toms} for details). The kernel of the quadratic operator $\bar \beta^{\mu}\beta^{\nu}\nabla_{\mu}\nabla_{\nu}$, whose eigenvalues are $\lambda_n^2$, can be written as
\bea
K(\tau; x, x') \equiv   \sum_{n=0}^{\infty}  e^{-i\tau \lambda_n^2}\, \Psi_{n}(x) \, \Psi_{n}^{\dagger}(x')\, ,
\eea
where $\tau$ plays the role of a regularization cutoff.
%
 With this, we can formally write 
\bea \label{nablaj}
\left< \nabla_{\mu}j_D^{\mu} \right> & = & 2  \lim_{\tau\to 0} {\rm Tr} [\beta_5 K(\tau; x, x)] \, ,
\eea
where the trace refers to $\Psi$ indices. The importance of the heat kernel regularization method relies in the asymptotic expansion of $K(\tau; x, x)$ in the limit $\tau\to0$
\be \label{Kexp} K(\tau; x, x)\sim -\frac{i}{16\pi^2\tau^2}\sum_{k=0}^{\infty} (i\tau)^k E_k(x)\, .\ee
The functions $E_k(x)$ are local geometric quantities, constructed from  the quadratic operator $\bar \beta^{\mu}\beta^{\nu}\nabla_{\mu}\nabla_{\nu}$; they depend on the metric and its first $2k$th derivatives. 
 The first few coefficients of the asymptotic kernel expansion are $E_0(x)=\mathbb I$, $E_1(x)=\frac{1}{6}R\, \mathbb I-\mathcal Q$, and 
\bea \label{E2}
E_2(x) & = & \left[\frac{1}{72}R^2-\frac{1}{180}R_{\mu\nu}R^{\mu\nu}+\frac{1}{180}R_{\alpha\beta\mu\nu}R^{\alpha\beta\mu\nu} \right] \mathbb I  \nonumber\\
 & - & \frac{1}{30} \Box R+\frac{1}{12}W_{\mu\nu}W^{\mu\nu}+\frac{1}{2}\mathcal Q^2-\frac{1}{6}R \mathcal Q+\frac{1}{6}\Box \mathcal Q,
\nonumber \eea
where $W_{\mu\nu}=[\nabla_{\mu}, \nabla_{\nu}]$,  
and $\mathcal Q$ is defined by writing the wave equation as
$-\bar \beta^{\mu}\beta^{\nu}\nabla_{\mu}\nabla_{\nu} \Psi\equiv (\Box+\mathcal Q)\Psi=0$. 
Explicit computations produce
\bea
\mathcal Q & \equiv &   \left( {\begin{array}{cc}
Q^{i}_{\hspace{0.15cm}j}  & 0  \\
0 & \bar Q^{i}_{\hspace{0.15cm}j} \\
  \end{array} } \right)\, ,
\eea
with $Q^{i}_{\hspace{0.15cm}j}=-[^+\Sigma^{\mu\nu}]^{i}_{\hspace{0.15cm}k}\, R_{\mu\nu\alpha\beta}\, [^+\Sigma^{\alpha\beta}]^{k}_{\hspace{0.15cm}j}$. Now, bringing the expansion (\ref{Kexp}) to Eq. (\ref{nablaj}) one finds that all terms in the sum (\ref{Kexp}) with $k>2$  clearly give a vanishing contribution in the $\tau\to 0$ limit. One can also check that for $k<2$ the terms in the sum vanish because the trace with $\beta_5$ selects the imaginary part, and  ${\rm Im}\{ {\rm Tr} \,   Q\}=0$. Henceforth,
\bea
\left< \nabla_{\mu}j_D^{\mu} \right> & = & \frac{i}{8\pi^2}{\rm Tr} [\beta_5 E_2(x)] \, .
\eea
The crucial point is then to evaluate this quantity. 
Using Eq. (\ref{E2}) one has ${\rm Tr}[ \beta_5 E_2] =\frac{1}{12} {\rm Tr }[\beta_5 W^{\mu\nu} W_{\mu\nu}]+\frac{1}{2} {\rm Tr } [\beta_5 \mathcal Q^2]$. Notice that the values of $W_{\mu\nu}$ and $\mathcal Q$ are related to the representations of the Lorentz group associated with the physical degrees of freedom of the electromagnetic theory, namely $(1,0)$ and $(0,1)$. This is in close analogy with the chiral anomaly for spin-$1/2$ fermions, where the $(1/2,0)$ and $(0,1/2)$ representations play an important role.
After a  long calculation, one  arrives at ${\rm Tr} [\beta_5 W^{\mu\nu} W_{\mu\nu}] = 2 i R^{\mu\nu\alpha\beta}\, {^{\star}R}_{\mu\nu \alpha\beta} $ and ${\rm Tr} (\beta_5 \mathcal Q^2)=-i  R^{\mu\nu\alpha\beta}\, ^{\star}R_{\mu\nu \alpha\beta} $ (details will be published elsewhere). Taking all factors into account, one gets
\bea \label{anomalyD}
\left< \nabla_{\mu}j_D^{\mu} \right>= \frac{1}{24 \pi^2}R_{\mu \nu \lambda \sigma} {^{\star}R} ^{\mu \nu \lambda \sigma}.
\eea
Since the heat kernel asymptotic series (\ref{Kexp}) does not depend on the vacuum state chosen, this expectation value is (vacuum) state independent. \\


{\bf 5. Conclusions and final comments}. 
The above result implies that the charge $Q_D$ associated with the duality symmetry of the Maxwell action  is no longer conserved in the quantum theory in a general spacetime; its time derivative is given by the spatial integral of Eq. (\ref{anomalyD}).  
Since in flat spacetime $Q_D$ represents the difference in  number between photons of opposite helicity \cite{Calkin}, this result can be interpreted as a nonconservation of the helicity of the quantum electromagnetic field in curved spacetimes. 


A physical  background where this anomaly may lead to observational consequences are rotating astrophysical objects, described approximately by a Kerr metric ($R_{\mu \nu \lambda \sigma} {^{\star}R} ^{\mu \nu \lambda \sigma}$ is proportional to the angular momentum of the source \cite{Ruffini}). Light rays coming from different sides of a rotating  object such as a black hole, galaxy, or cluster, not only would bend around, but  an effective difference in polarization could also be induced between them. 
In particular, this effect would affect the polarization of the cosmic microwave background photons. The quantitative details for phenomenological implications will be analyzed in a future work.

Interestingly, the anomaly (\ref{anomalyD}) can be understood as a physical realization of the   Hirzebruch signature (index) theorem \cite{geometry}. 
The anomaly arises as the difference in the number of right-handed  and left-handed zero-eigenvalue solutions of the operator $\beta^{\mu}\nabla_{\mu}$,  
 $\int d^4x \sqrt{-g}\left< \nabla_{\mu}j_D^{\mu} \right>=2[n_L-n_R]$. $n_L$ and $n_R$ can be computed from the $(0,1)$ and $(1,0)$ irreducible representations of the Lorentz group \cite{christensen-duff}, respectively, and one obtains  agreement with Eq. (\ref{anomalyD}).  This is also in analogy with the fermionic chiral anomaly, which can also be obtained from an index associated with the $(1/2,0)$ and $(0,1/2)$ irreducible representations of the Lorentz group.\\


{\it Acknowledgments}.   
This work was  supported  by the  Grants. No.\ FIS2014-57387-C3-1-P, No.\  CPANPHY-1205388, and No.\ MPNS of COST Action No. MP1210,  the Severo Ochoa program SEV-2014-0398 and NSF Grants No. PHY-1403943 and No. PHY-1552603. A. d. R. is supported by the FPU Ph.D. fellowship FPU13/04948 and FPU research stay Grant EST14/00587.
I.A. thanks E. Mottola, R. Wald, and R. Gambini for  discussions. A. d. R. is grateful to  the members of the Gravity Theory Group of Louisiana State University  for their hospitality during his stay there.  


\begin{thebibliography}{99}

\nonfrenchspacing


\bibitem{deser-teitelboim} S.\ Deser and C.\ Teitelboim, {\it Phys.\ Rev.\ D}{\bf 13}, 1592 (1976). 
S. Deser, {\it J. Phys. A} {\bf 15}, 1053 (1982).

\bibitem{ALN14} I. Agullo, A. Landete and J. Navarro-Salas, {\it Phys. Rev. D} {\bf 90}, 124067 (2014).

\bibitem{DKZ87} A. D. Dolgov, I. B. Khriplovich, A. I. Vainshtein,  V. I. Zakharov, {\it Nucl. Phys. B} {\bf 315}, 138 (1989). 




\bibitem{ABJ} S. L. Adler,  {\it Phys. Rev.} {\bf 177}, 2426 (1969). J. S. Bell and R. Jackiw, {\it Nuovo Cimento A} {\bf 60}, 47 (1969). 

\bibitem{Kimura} T. Kimura {\it Prog. Theor. Phys. } {\bf 42}, 1191 (1969). 







\bibitem{Fujikawa} K. Fujikawa, {\it Phys. Rev. Lett.} {\bf 42}, 1195 (1979); {\it Phys. Rev. D} {\bf 21}, 2848 (1980).

\bibitem{Fbook} K. Fujikawa and H. Suzuki, {\it Path Integrals and Quantum Anomalies}, Oxford University Press, Oxford (2004).

\bibitem{Calkin} M. G. Calkin, {\it Am. J. Phys.} {\bf 33}, 958 (1965).

\bibitem{z0}Notice that  the ``zero-component" $Z_{0}$ remains an arbitrary gauge function, and can be chosen to be zero in such a way that  $A_0$ does not transform under duality \cite{deser-teitelboim}. This is convenient since in Maxwell theory $A_0$ is a nondynamical variable, i.e. another gauge function.

\bibitem{Berger1999} M. A. Berger, {\it Plasma Phys. Control. Fusion} {\bf 41} B167-B175 (1999) .
\bibitem{Weinberg64} S. Weinberg,  {\it Phys. Rev.} {\bf 134}, B882 (1964).

\bibitem{Dowker}  J.S. Dowker and Y. P. Dowker {\it Proc. R. Soc. A.} {\bf  294} 175 (1966); J.S. Dowker, {\it J.Phys. A: Math. Gen.}  {\bf 11}, 2 (1978).

\bibitem{Geroch} R. Geroch,  {\it Handbook of spacetime. Chapter 15: Spinors}, Eds. A. Asthekar and V. Petkov, Springer (2015); J. Wess, J. Bagger, {\it Supersymmetry and Supergravity} Princeton University Press, (1992). 


\bibitem{footnote}  
There is another way to understand why Maxwell's equations can be written as first-order differential equations for the potentials. Maxwell's equations can be  recovered from the following two conditions: $(i)$  $d\, ^{+}F=0$, and $(ii)$ $^{+}F_{ab}$ being self-dual, i.e., $^{+}F_{ab}=\frac{1}{2}\left[F_{ab}+i{^{\star}F}_{ab} \right]$. The equation $d\, ^{+}F=0$ allows one to define the potential $A_+$; while the self-duality property   ${^{+}F}_{ab}=\frac{ i}{2}\epsilon_{abcd}{^{+}F}^{cd}$, when written in terms of $A_+$, gives the dynamical equations (\ref{alphaMaxwellA}) for $A_+$. 


\bibitem{photon} 
The lower component in this field $\Psi$ is related to the upper one by charge conjugation. In contrast to the Dirac case,  this operation just consists of complex conjugation. 
This simplicity in the structure of $\Psi$ can be heuristically regarded  as a manifestation of the fact that the photon is its own antiparticle.


\bibitem{constraints} R. Jackiw, {\it Diverse Topics in Theoretical and Mathematical Physics}, 367-381, (World Scientific, Singapore, 1995); D. J. Toms, { Phys. Rev. D} {\bf 92}, 105026 (2015).



















\bibitem{parker-toms}L. Parker and D.J. Toms, {\it Quantum Field Theory in Curved Spacetime: Quantized Fields
and Gravity} (Cambridge University Press, Cambridge, England, 2009).

\bibitem{reuter} M. Reuter, {\it Phys. Rev. D} {\bf 37}, 1456 (1988).

	












\bibitem{Ruffini} C. Cherubini et al. {\it Int. J. Mod. Phys. D} {\bf 11} (2002) 827.

\bibitem{geometry} T. Eguchi, P. B. Gilkey, A. J. Hanson, {\it Gravitation, Gauge Theories, and Differential Geometry}, Phys. Rep. C {\bf 66}, 213-393 (1980); M. Nakahara, {\it Geometry, Topology, and Physics. Second Edition}, IOP Publishing, Bristol and Philadelphia (2003).

\bibitem{christensen-duff} S. M. Christensen, M.J. Duff, {\it Nucl. Phys. B} {\bf 154} (1979) 301-342; {\it Phys. Lett.} {\bf 76B}, 571 (1978).


\newpage



























\end{thebibliography}
\end{document}